\documentclass[draftclsnofoot,onecolumn, 12pt]{IEEEtran}
\usepackage{color}
\usepackage{graphicx, pstool}
\usepackage{amsmath}
\usepackage{amsfonts}
\usepackage{amssymb}
\usepackage{booktabs}
\usepackage{dsfont}
\usepackage{cases}
\usepackage{tabularx}
\usepackage{boxedminipage}
\usepackage{url}
\usepackage{mathrsfs}
\usepackage{cite}
\usepackage{authblk}
\usepackage{amsmath}
\usepackage{setspace}
\usepackage{mathtools}
\usepackage{breqn}
\newtheorem{theorem}{Theorem}
\newtheorem{remark}{Remark}
\newtheorem{lemma}{Lemma}

\begin{document}

\title{ Buffer-Aided Relaying For The Two-Hop Full-Duplex Relay Channel With Self-Interference}
\author{Mohsen Mohammadkhani Razlighi and Nikola Zlatanov
\thanks{M. M. Razlighi and N. Zlatanov are  with the Department of Electrical and Computer Systems Engineering, Monash University, Melbourne, VIC 3800, Australia (e-mails: mohsen.mohammadkhanirazlighi@monash.edu and nikola.zlatanov@monash.edu).}
}
\maketitle

\begin{abstract}
In this paper, we investigate  the fading two-hop full-duplex (FD) relay channel with self-interference, which is comprised of a source, an FD relay impaired by self-interference, and a destination, where a direct source-destination link does not exist. For this channel, we propose three buffer-aided relaying schemes with adaptive reception-transmission at the FD relay  for the cases when the source and the relay both perform adaptive-rate transmission with adaptive-power allocation, adaptive-rate transmission with fixed-power allocation, and fixed-rate transmission, respectively. The  proposed buffer-aided relaying schemes   enable  the FD relay to adaptively select to either receive, transmit,  or simultaneously receive and transmit  in a given time slot based on the qualities of the receiving,  transmitting, and self-interference channels; a degree-of-freedom unavailable without buffer-aided relaying.  Our numerical results show that significant performance gains are achieved using the proposed buffer-aided relaying schemes compared to   conventional FD relaying, where the FD relay is forced to always simultaneously receive and transmit, and to   buffer-aided half-duplex relaying, where the half-duplex relay  cannot simultaneously receive and transmit. The main implication of this work is  that FD relaying systems without buffer-aided relaying miss-out on significant performance gains.
\end{abstract}

\section{Introduction}\label{Sec-Intro}
Relays play an important role in  wireless communications for increasing the data rate  between a source and a destination  \cite{cover}. 
In general, the relay can operate in two different modes, namely,  full-duplex (FD) mode and   half-duplex (HD) mode. In the FD mode, transmission and reception at the FD relay can occur simultaneously and in the same frequency band. However, due to the in-band simultaneous reception and transmission, FD relays are impaired by self-interference (SI), which occurs due to leakage of energy from the transmitter-end into the receiver-end of the FD relay. Currently, there are advanced hardware designs which can suppress the SI by about 110 dB in certain scenarios, see \cite{7024120}. Because of this, FD relaying with SI is gaining considerable research interest \cite{7024120,7051286}. On the other hand, in the HD mode, transmission and reception take place in the same frequency band but in different time slots, or in the same time slot but in different frequency bands. As a result, HD relays avoid the creation of SI. However, since an FD relay  uses twice the resources compared to an HD relay, the achievable data rates of an FD relaying system may be significantly higher than that of an HD relaying system.

One of the first immediate applications of   FD relaying is expected to be in providing support to HD base stations. In particular, the idea is to deploy FD relays around HD base stations, which will relay information from the HD base stations  to users that are at significant distances from the base stations. The  system model resulting from such a scenario is the two-hop FD relay channel, which  is comprised of a source, an FD relay, and a destination, where a direct source-destination link does not exist  due to the assumed large distance between the source and the destination. In this paper, we will investigate new achievable rates/throughputs for this system model, i.e, for the two-hop FD relay channel with SI and links impaired by fading.

 The two-hop relay channel with and without fading has been extensively investigated in the literature both for HD relaying as well as FD relaying with and without SI. In particular, the capacity of the two-hop HD relay channel without fading was derived in \cite{zlatanov2014capacity-globecom}. On the other hand, only achievable rates are known for the two-hop HD relay channel with fading. Specifically, \cite{1435648} proposed a conventional decode-and-forward (DF) relaying scheme, where the HD relay switches between  reception and transmission in a prefixed manner. On the other hand, \cite{BA-relaying-adaptive-rate}  proposed a buffer-aided relaying scheme  where, in each time slot, the HD relay selects to either receive or transmit  based on the qualities of the receiving and transmitting channels. As a result, the   rate achieved by the scheme in \cite{BA-relaying-adaptive-rate}  is larger than the   rate achieved by the scheme in \cite{1435648}, showing that buffers improve the performance of HD relays. The capacity of the two-hop FD relay channel with an idealized FD relay without SI was derived in \cite{cover} and \cite{1435648} for the cases with and without fading, respectively. Recently, the capacity of the Gaussian two-hop FD relay channel with SI and without fading was derived in \cite{zlatanov_fd_cap}. However, for the two-hop FD relay channel with SI and fading only achievable rates are known for certain special cases, such as an SI channel without fading, see \cite{7370810}.

Motivated by the lack of advanced schemes for the general two-hop FD relay channel with SI and fading, in this paper, we investigate this channel and propose novel achievable rates/throughputs. The novel   rates/throughputs are achieved using buffer-aided relaying. Thereby,  similar to HD relays, we show that  buffers also improve the performance of FD relays with SI. This means that buffer-aided relaying should become an integral part of FD relaying systems, i.e., that FD relaying systems without buffer-aided relaying miss-out on significant performance gains.

The proposed novel buffer-aided relaying schemes for the two-hop FD relay channel with SI and fading enable the FD relay to  \textit{select adaptively  either to receive, transmit,  or simultaneously receive and transmit  in a given time slot} based on the qualities of the receiving,  transmitting, and SI channels such that the achievable data rate/throughput is maximized. Note that such a degree of freedom is not available if a buffer is not employed. Specifically, we propose three buffer-aided relaying schemes with adaptive reception-transmission at the FD relay, for the cases when both the source and the relay perform adaptive-rate transmission with adaptive-power allocation, adaptive-rate transmission with fixed-power allocation, and fixed-rate transmission, respectively.  The proposed buffer-aided schemes significantly improve the achievable rate/throughput of the considered relay channel compared to existing schemes. In particular,  our numerical results show that significant performance gains are achieved using the proposed buffer-aided relaying schemes compared to    conventional FD relaying, where the FD relay is forced to always simultaneously receive and transmit, and to  buffer-aided HD relaying, where the   HD relay cannot simultaneously receive and transmit. 

 We note that  buffer-aided relaying schemes were also proposed in \cite{7560604,7523270,7317764,7370810,6510556,7067380,7478633,Taghizadeh2016266,7350164,7480823,7145471,7120945,7039198,7036071,6177989,7765086,7374682,7317590,7114345,7031378} and references therein. Most of these works investigate HD buffer-aided relaying systems. FD buffer-aided systems were investigated in \cite{7560604} and \cite{7523270}. However, these works differ from our work since \cite{7560604}  assumes that the SI at the FD relay is negligeable,  which may not be a realistic model for all scenarios in practice. On the other hand, \cite{7523270} assumes that the SI channel is fixed and does not vary with time, which also may not be an accurate model of the SI channel. In particular, in   practical wireless communications, due to the movement of objects/reflections, the SI channel also varies with time. Hence, contrary to \cite{7523270},   in this paper, the   SI channel is assumed to vary with time. In addition, we investigate the case when both the source and the relay transmit with a fixed rate  in all time slots; a scenario not investigated in \cite{7523270}.

 This paper is organized as follows. In Section \ref{Sec-Sys}, we present the system and channel models. In Section \ref{Sec-BAR}, we formulate a general FD buffer-aided relaying scheme with adaptive reception-transmission at the FD relay. In Section \ref{Sec-BAART}, we propose specific schemes for the cases when both the source and the relay adapt their transmission rates with adaptive- and fixed-power allocation, and derive the corresponding optimal buffer-aided relaying schemes. The throughput of the considered relay channel with fixed-rate transmission is derived in Section \ref{Sec-FR}. Simulation and numerical results are provided in Section \ref{Sec-Num}, and the conclusions are drawn in Section \ref{Sec-Conc}.

\section{System and Channel Models}\label{Sec-Sys}
\begin{figure}[t]
\centering\includegraphics[width=3.5in]{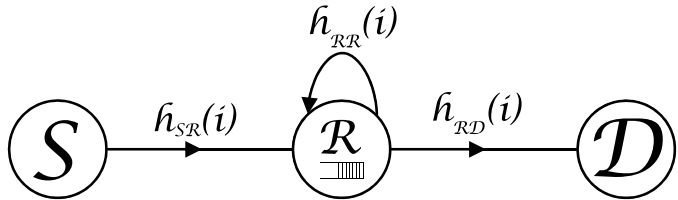}
\caption{Relaying system, comprised of a source, S, an FD relay impaired by SI, R, and a destination, D, where a direct S-D link is not available.}
\label{Sys_model_fig}
\end{figure}

We consider the two-hop FD relay channel, which is comprised of a source, S, an FD relay impaired by SI, R, and a destination, D, where a direct S-D link does not exist, cf. Fig.~\ref{Sys_model_fig}. 
In addition, we assume that the FD relay is a DF relay equipped with a sufficiently large buffer in which it can store incoming data and from which it can extract data for transmission to the destination.  

\subsection{Channel Model}
We assume that the S-R and R-D links are complex-valued additive white Gaussian noise (AWGN) channels impaired by slow fading. Furthermore, similar to the majority of related papers\cite{5985554,Day}, we also assume that the SI channel is impaired by slow fading. Thereby, the SI channel also varies with time. We assume that the transmission time is divided into $ N\to\infty $ time slots. Furthermore, we assume   that the fading is constant during one time slot and changes from one time slot to the next. Let $h_{SR}(i)$ and $h_{RD}(i)$ denote the complex-valued fading gains of the S-R and R-D channels in time slot $i$, respectively, and let $h_{RR}(i)$ denote the complex-valued fading gain of the SI channel in time slot $i$. Moreover, let $\sigma _{{n_{R}}}^2$ and $\sigma _{{n_{D}}}^2$ denote the variances of the complex-valued AWGNs at the relay and the destination, respectively. For convenience, and without loss of generality, we   define normalized squared fading gains for the S-R, R-D, and SI channels as $\gamma_{SR}(i)=|h_{SR}(i)|^2/\sigma _{{n_{R}}}^2$, $\gamma_{RD}(i)=|h_{RD}(i)|^2/\sigma _{{n_{D}}}^2$, and $\gamma_{RR}(i)=|h_{RR}(i)|^2/\sigma _{{n_{R}}}^2$, respectively.  Let ${P_S}(i)$ and ${P_R}(i)$ denote the transmit powers of the source and the relay in time slot $i$, respectively. We assume that the SI, which is received via the SI channel in any symbol interval of time slot $i$, is independent and identically distributed  (i.i.d.)  according to the zero-mean Gaussian distribution with variance $P_R(i) |h_{RR}(i)|^2$, an assumption similar to the majority of related works \cite{5985554,Day,6353396,Bharadia}. This assumption is realistic due to the combined effect of various sources of imperfections in the SI cancellation process, and can also be considered as the worst-case scenario for SI \cite{6283743}. As a result, in time slot $i$, the S-R channel is a complex-valued AWGN channel with channel gain $h_{SR}(i)$ and noise variance $P_R(i) |h_{RR}(i)|^2+ \sigma _{{n_{R}}}^2$. Hence, the capacity of this channel in time slot $i$ is obtained as  
\begin{equation}
{C_{SR}}(i)={\log _2}\left( {1 + \frac{{{P_S}(i){\gamma _{SR}}(i)}}{{{P_R}(i){\gamma _{RR}}(i) + 1}}} \right).\
 \end{equation}
On the other hand, in time slot $i$, the R-D channel is also a complex-valued AWGN channel with channel gain $h_{RD}(i)$ and noise variance $\sigma _{{n_{D}}}^2$. Hence, the capacity of this channel in time slot $i$ is obtained as  
\begin{equation}
{C_{RD}}(i)={\log _2}\left(1 + {P_R}(i){\gamma _{RD}}(i)\right).\
\end{equation}
 In time slot $i$, we assume that the source and the relay transmit  codewords encoded with a capacity achieving code, i.e., codewords  comprised of $n\to\infty$ symbols that are generated independently from  complex-valued zero-mean Gaussian distributions with variances ${P_S}(i)$ and ${P_R}(i)$, respectively. The data rates of the codewords transmitted from the source and the relay in time slot $i$ are denoted by $R_{SR}(i)$ and $R_{RD}(i)$, respectively. The value of $R_{SR}(i)$ and $R_{RD}(i)$ will be defined  later on. 
 
\section{General FD Buffer-Aided Relying} \label{Sec-BAR}
In this section, we formulate a general FD buffer-aided relaying scheme with adaptive reception-transmission at the FD relay.

\subsection{Problem Formulation}

Depending on whether the source and/or the relay are silent, we have four different states  for the considered two-hop FD relay channel with SI in each time slot:
\begin{itemize}\label{items_state}
\item \textit{State 0:} S and R are both silent.
\item \textit{State 1:} S transmits and R receives without transmitting.
\item \textit{State 2:} R transmits and S is silent.
\item \textit{State 3:} S transmits and R simultaneously receives and transmits.
\end{itemize}
To model these four states, we define three binary variables for time slot $i$, $q_{1}(i), q_{2}(i)$, and $q_{3}(i)$, as
\begin{align}
q_1(i) =& \left\{ 
\begin{array}{ll}
1& \textrm{if S transmits and R receives without transmitting in time slot $i$}\\
0& \textrm{otherwise},
\end{array}
 \right.\\
q_2(i) =& \left\{ 
\begin{array}{ll}
1& \textrm{if R transmits and S is silent in time slot $i$}\\
0& \textrm{otherwise},
\end{array}
 \right.\\ 
q_3(i) =& \left\{ 
\begin{array}{ll}
1& \textrm{if S transmits and R simultaneously receives and transmits in time slot $i$}\\
0& \textrm{otherwise}.
\end{array}
 \right.
\end{align}
Since the two-hop FD relay channel can be in one and only one of the four states in time slot $i$, the following has to hold 
\begin{align}\label{eq_1}
q_{1}(i)+q_{2}(i)+q_{3}(i)\in\{0,1\},
\end{align}
  where if $q_{1}(i)+q_{2}(i)+q_{3}(i)=0$ occurs, it means that the system is in State 0, i.e., both S and R are silent in time slot $i$. 

To generalize the FD buffer-aided  relaying scheme even further, we assume different  transmit powers at the source and the relay during the different states. In particular, let  $P_S^{(1)}(i)$ and $P_S^{(3)}(i)$ denote  the powers of the source when $q_1(i)=1$ and $q_3(i)=1$, respectively, and   let  $P_R^{(2)}(i)$ and $P_R^{(3)}(i)$ denote  the powers of the relay when $q_2(i)=1$ and $q_3(i)=1$, respectively. Obviously,  $P_S^{(1)}(i)=0$, $P_R^{(2)}(i)=0$, and $P_S^{(3)}(i)=P_R^{(3)}(i)=0$ when $q_1(i)=0$, $q_2(i)=0$, and $q_3(i)=0$ hold, respectively. 

In the following sections, we provide the optimal values for the state selection variables $q_k(i)$, $\forall k$, which maximize the achievable rate and the throughput of the considered relay channel. To this end, we define the following auxiliary   optimal state selection scheme
\begin{equation}\label{eq_scheme_AP}
\textrm{Optimal Scheme}  = \left\{ {\begin{array}{ll}
q_1(i)=1&{\textrm{if}\;\Lambda_1(i) }> \Lambda_3(i) \\
& {\textrm{ and }} \Lambda_1\geq \Lambda_2(i) \\
q_2(i)=1&{\textrm{if}\;\Lambda_2(i) }> \Lambda_3(i)\\
& {\textrm{ and }} \Lambda_2>\Lambda_1(i) \\
q_3(i)=1&{\textrm{if}\;\Lambda_3(i) }\geq \Lambda_1(i)\\
& {\textrm{ and }} \Lambda_3\geq \Lambda_2(i),
\end{array}} \right. 
\end{equation}
where $\Lambda_1(i)$, $\Lambda_2(i)$,   and $\Lambda_3(i)$  will be defined later on, cf. Theorems~\ref{Theo_AP} -~\ref{Theo_FR}.

\section{Buffer-Aided Relaying with Adaptive-Rate Transmission}\label{Sec-BAART}
In this section, we provide  buffer-aided relaying schemes for the case when both the source and the relay adapt their transmission rates to the underlaying channels in each time slot. Thereby, we provide two buffer-aided relaying schemes with adaptive-rate transmission; one in which both the source and the relay also adapt their transmit powers to the underlying channels in each time slot, and the other one in which both the source and the relay transmit with fixed-powers in each time slot. 
\subsection{Problem Formulation for Buffer-Aided Relaying with Adaptive-Rate Transmission}
Using the state selection variables $q_k(i)$, $\forall k$, and the transmit powers of the source and the relay for each possible state, we can write the capacities of the S-R and R-D channels in time slot $i$, $C_{SR}(i)$ and $C_{RD}(i)$, as
\begin{align}
C_{SR}(i)= &q_1(i) {\log _2}\left(1 + P_S^{(1)}(i){\gamma _{SR}}(i)\right)
+ q_3(i){\log _2}\left( {1 + \frac{{P_S^{(3)}(i) {\gamma _{SR}}(i)}}{{P_R^{(3)}(i) {\gamma _{RR}}(i) + 1}}} \right),\label{eq_2a}\\
C_{RD}(i) =& q_2(i) {\log _2}\left(1 + P_R^{(2)}(i){\gamma _{RD}}(i)\right)
+ q_3(i) {\log _2}\left(1 + P_R^{(3)}(i){\gamma _{RD}}(i)\right). \qquad\quad\label{eq_2b}
\end{align}

Now, since the source is assumed to be backlogged, we can set  the transmission rate at the source   in time slot $i$, $R_{SR}(i)$,  to $R_{SR}(i)=C_{SR}(i)$, where $C_{SR}(i)$ is given by (\ref{eq_2a}). As a result, the achievable  rate  on the S-R   channel during $N\to\infty$ time slots, denoted by $\bar R_{SR}$, is obtained as
\begin{align}
&\bar R_{SR}  = \lim_{N\to\infty}\frac 1N \sum_{i=1}^N R_{SR}(i) \label{eq_3a}\\
&= \lim_{N\to\infty}\frac 1N \sum_{i=1}^N \left[ q_1(i) {\log _2}\left(1 + P_S^{(1)}(i){\gamma _{SR}}(i)\right)+ q_3(i){\log _2}\left( {1 + \frac{{P_S^{(3)}(i) {\gamma _{SR}}(i)}}{{P_R^{(3)}(i) {\gamma _{RR}}(i) + 1}}} \right) \right].\nonumber 
\end{align}
On the other hand, the relay can transmit only if it has information stored in its buffer. Let $Q(i)$, denote the amount of (normalized) information in bits/symbol in the buffer of the relay at the end (beginning) of time slot $i$ (time slot $i+1$). Then, we can  set  the transmission rate at the relay   in time slot $i$, $R_{RD}(i)$,  to $R_{RD}(i)=\min\{Q(i-1),C_{RD}(i)\}$, where $C_{RD}(i)$ is given by (\ref{eq_2b}). As a result, the achievable  rate  on the R-D   channel during $N\to\infty$ time slots, denoted by $\bar R_{RD}$, is obtained as
\begin{align}
\bar R_{RD} &= \lim_{N\to\infty}\frac 1N \sum_{i=1}^N R_{RD}(i) \nonumber\\
&= \lim_{N\to\infty}\hspace{-0.5mm}\frac 1N\hspace{-0.5mm} \sum_{i=1}^N\hspace{-1mm} \Big[ \min\Big\{Q(i-1),q_2(i) {\log _2}\left(1+ P_R^{(2)}(i){\gamma _{RD}}(i)\right) \nonumber\\
&+ q_3(i) {\log _2}\left(1 + P_R^{(3)}(i){\gamma _{RD}}(i)\right)\Big\}\Big],\label{eq_3b}
\end{align} 
where $Q(i)$ is obtained  recursively  as 
\begin{align}\label{eq_3c}
Q(i)=Q(i-1)+R_{SR}(i)-R_{RD}(i).
\end{align}
  
Our task in this section is to maximize the achievable rate of the considered relay channel given by (\ref{eq_3b}). To this end, we use the following Lemma from    \cite[Theorem 1]{BA-relaying-adaptive-rate}.
\begin{lemma}\label{lemma_1}
The  data rate extracted from the buffer of the relay and transmitted to the destination during $N\to\infty$ time slots, $\bar R_{RD}$, is maximized when the following condition holds 
\begin{align}\label{eq_AD_FP} 
&\lim_{N\to\infty}\frac 1N \sum_{i=1}^N \left[ q_1(i) {\log _2}\left(1 + P_S^{(1)}(i){\gamma _{SR}}(i)\right) 
+ q_3(i){\log _2}\left( {1 + \frac{{P_S^{(3)}(i) {\gamma _{SR}}(i)}}{{P_R^{(3)}(i) {\gamma _{RR}}(i) + 1}}} \right) \right]
\nonumber\\
&=\lim_{N\to\infty}\frac 1N \sum_{i=1}^N \left[ q_2(i) {\log _2}\left(1 + P_R^{(2)}(i){\gamma _{RD}}(i)\right)
+ q_3(i) {\log _2}\left(1 + P_R^{(3)}(i){\gamma _{RD}}(i)\right)\right].
\end{align}
Moreover, when condition (\ref{eq_AD_FP}) holds, the rate, $\bar R_{RD}$, given by (\ref{eq_3b}), simplifies to  
\begin{dmath}
\bar R_{RD}   = \lim_{N\to\infty}\frac 1N \sum_{i=1}^N \left[  q_2(i) {\log _2}\left(1 + P_R^{(2)}(i){\gamma _{RD}}(i)\right) + q_3(i) {\log _2}\left(1 + P_R^{(3)}(i){\gamma _{RD}}(i)\right) \right]\label{eq_4a}.
\end{dmath} 
\end{lemma}
\begin{IEEEproof}
See \cite[Theorem 1]{BA-relaying-adaptive-rate} for the proof.
\end{IEEEproof}

Lemma~\ref{lemma_1} is very convenient since it provides an expression for the maximum data rate, $\bar R_{RD}$, which is independent of the state of the buffer $Q(i)$. This is because, when  condition  (\ref{eq_AD_FP}) holds, the number of time slots for  which $R_{RD}(i)=\min\{Q(i-1),C_{RD}(i)\}=Q(i-1)$ occurs is negligible compared to the number of time slots for  which $R_{RD}(i)=\min\{Q(i-1),C_{RD}(i)\}=C_{RD}(i)$ occurs when $N\to\infty$, see \cite{BA-relaying-adaptive-rate}. In other words, when condition  (\ref{eq_AD_FP}) holds, we can consider that the buffer at the relay has enough information almost always.

\subsection{Buffer-Aided Relaying with Adaptive-Rate Transmission and Adaptive-Power Allocation}\label{Sec-AP}
In this subsection, we assume that the source and the relay can also adapt their transmit powers in each time slot such that a long-term average power constraint $P$ is satisfied. More precisely, $P_S^{(1)}(i)$, $P_S^{(3)}(i)$, $P_R^{(2)}(i)$, and $P_R^{(3)}(i)$ have to satisfy the following constraint 
\begin{align}\label{eq_n1}
&\mathop {\lim }\limits_{N \to \infty } \frac{1}{N}\sum\limits_{i = 1}^N \left[ q_1(i) P_S^{(1)}(i)+ q_2(i) P_S^{(3)}(i)\right]\nonumber\\
&+\mathop {\lim }\limits_{N \to \infty } \frac{1}{N}\sum\limits_{i = 1}^N  \left[ q_2(i) P_R^{(2)}(i) + q_3(i) P_R^{(3)}(i) \right]\le P. 
\end{align}
Now, employing Lemma~\ref{lemma_1}, we devise the following optimization problem for maximizing the  rate, $\bar R_{RD}$, when $N\to\infty$
\\
\vspace{-1mm}
\begingroup
\addtolength{\jot}{-0.5em}
\begin{align}
& {\underset{q_k(i),P_S^{(k)}(i),P_R^{(k)}(i),\forall i,k} {  \textrm{Maximize:} }}
 \frac 1N \sum_{i=1}^N \left[  q_2(i) {\log _2}\left(1 + P_R^{(2)}(i){\gamma _{RD}}(i)\right)
+ q_3(i) {\log _2}\left(1 + P_R^{(3)}(i){\gamma _{RD}}(i)\right) \right] \nonumber\\
&{\rm{Subject\;\;  to \; :}}  \nonumber\\
&\qquad\qquad{\rm C1:}  \frac {1}{N} \sum_{i=1}^N \left[ q_1(i) {\log _2}\left(1 + P_S^{(1)}(i){\gamma _{SR}}(i)\right)+ q_3(i){\log _2}\left( {1 + \frac{{P_S^{(3)}(i) {\gamma _{SR}}(i)}}{{P_R^{(3)}(i) {\gamma _{RR}}(i) + 1}}} \right) \right] \nonumber\\
\rm & \qquad\qquad\quad= \frac 1N \sum_{i=1}^N \left[  q_2(i) {\log _2}\left(1 + P_R^{(2)}(i){\gamma _{RD}}(i)\right) + q_3(i) {\log _2}\left(1 + P_R^{(3)}(i){\gamma _{RD}}(i)\right) \right]\nonumber\\[0.1em]
& \qquad\qquad{\rm C2:} q_k(i)\in\{0,1\}, \textrm{ for } k=1,2,3  \nonumber\\
& \qquad\qquad{\rm C3:} q_1(i)+q_2(i)+q_3(i)\in\{0,1\}\nonumber\\
& \qquad\qquad{\rm C4:}\frac{1}{N}\sum\limits_{i = 1}^N \left[ q_1(i) P_S^{(1)}(i)+ q_2(i) P_S^{(3)}(i)\right]+ \frac{1}{N}\sum\limits_{i = 1}^N  \left[ q_2(i) P_R^{(2)}(i) + q_3(i) P_R^{(3)}(i) \right]\le P \nonumber\\[0.15em]
&\qquad\qquad{\rm C5:} P_S^{(k)}(i) \ge 0, P_R^{(k)}(i) \ge 0, \forall k,\label{eq_op_AP}
\end{align}
\endgroup
where constraint C1 ensures that (\ref{eq_AD_FP}) holds, C2 constrains  the values that $q_k(i)$, $\forall k,i$, can assume, C3 ensures no more than one state is active at a given time slot $i$,   C4 ensures the joint source-relay power constraint in (\ref{eq_n1}) holds, and constraint C5 ensure  that the transmit powers are non-negative. 
\vspace{-0.1mm}

In the following theorem, we provide the solution of problem (\ref{eq_op_AP}), which maximizes the  achievable rate of the considered buffer-aided FD relay channel with SI for adaptive-rate transmission with adaptive-power allocation.

\vspace{-0.1mm}
\begin{theorem}\label{Theo_AP}
The optimal  state selection variables $q_k(i)$, $\forall k,i$, found as the solution of (\ref{eq_op_AP}),  are given in (\ref{eq_scheme_AP}),
where  $\Lambda_1(i)$, $\Lambda_2(i)$,   and $\Lambda_3(i)$ are defined   as 
\begin{align}
\Lambda_1(i) &= \mu \log_2\left(1 + {P_S^{(1)}(i)}{\gamma _{SR}}(i)\right) - \zeta {P_S^{(1)}(i)},\label{eq_AP_S1}\\
\Lambda_2(i)&= (1 - \mu )\log_2\left(1 + {P_R^{(2)}(i)}{\gamma _{RD}}(i)\right) - \zeta {P_R^{(2)}(i)},\label{eq_AP_R1}\\
\Lambda_3(i)&= \mu \log_2\left(1 + \frac{{{P_S^{(3)}(i)}{\gamma _{SR}}(i)}}{{{P_R^{(3)}(i)}{\gamma _{RR}}(i) + 1}}\right) - \zeta {P_S^{(3)}(i)}\nonumber\\
&+ (1 - \mu )\log_2\left(1 + {P_R^{(3)}(i)}{\gamma _{RD}}(i)\right) - \zeta {P_R^{(3)}(i)}.\label{eq_AP_R2}
\end{align}
Moreover, the optimal $P_S^{(1)}(i)$ and $P_R^{(2)}(i)$, found as the solution of (\ref{eq_op_AP}), are given by
\vspace{-1mm}
\begin{eqnarray}\label{eq_scheme_AP_PS1}
{P_S^{(1)}(i)} &= \left\{ {\begin{array}{*{20}{l}}
{\begin{array}{*{20}{c}}
{\frac{\rho }{\eta  } - \frac{1}{{{\gamma _{SR}}(i)}}}\\
0
\end{array}}&{\begin{array}{*{20}{l}}
{\textrm{if}\;{\gamma _{SR}}(i) > \eta  /\rho }\\
\textrm{otherwise,}
\end{array}}
\end{array}} \right. \\
{P_R^{(2)}(i)} &= \left\{ {\begin{array}{*{20}{l}}\label{eq_scheme_AP_PR2}
{\begin{array}{*{20}{c}}
{\frac{1}{\eta  } - \frac{1}{{{\eta  _{RD}}}(i)}}\;\;\;\;\\
0
\end{array}}&{\begin{array}{*{20}{l}}
{\textrm{if}\;{\gamma _{RD}}(i) > \eta  }\\
\textrm{otherwise,}
\end{array}}
\end{array}} \right. \
\end{eqnarray}
 where  $\eta \triangleq\frac{\zeta \ln(2)}{1-\mu}$ and $\rho\triangleq\frac{\mu}{1-\mu}$. Whereas, the optimal $P_S^{(3)}(i)$ and $P_R^{(3)}(i)$ are obtained as the solution of the following system of two equations 
\begin{align}
& \frac{{-\mu {\gamma _{RR}}(i){\gamma _{SR}}(i){P_S^{(3)}(i)}}}{{{\mkern 1mu} (1 + {P_R^{(3)}(i)}{\gamma _{RR}}(i)){\mkern 1mu} (1 + {P_R^{(3)}(i)}{\gamma _{RR}}(i) + {P_S^{(3)}(i)}{\gamma _{SR}}(i))}}\nonumber\\
&+\frac{{(1 - \mu ){\gamma _{RD}}(i)}}{{{\mkern 1mu} (1 + {P_R^{(3)}(i)}{\gamma _{RD}}(i))}} - \ln(2)\zeta= 0,\label{eq_AP_PR31}
\end{align}
\begin{align}
&\frac{{\mu{\gamma _{SR}}(i)}}{{{\mkern 1mu} (1 + {P_R^{(3)}(i)}{\gamma _{RR}}(i) + {P_S^{(3)}(i)}{\gamma _{SR}}(i))}} - \ln(2)\zeta = 0.\label{eq_AP_PR32}
\end{align}
In (\ref{eq_AP_PR31})-(\ref{eq_AP_PR32}), $\mu$ and $\zeta$ are constants found such that C1 and C4 in (\ref{eq_op_AP}) are satisfied, respectively.
 \end{theorem} 
\begin{IEEEproof}
Please refer to Appendix~\ref{app_AP} for the proof.
\end{IEEEproof}

\vspace{-1mm}
\subsection{Buffer-Aided Relaying  with Adaptive-Rate Transmission and Fixed-Power Allocation}\label{Sec-FP}
In this subsection, we assume that the powers at the source and the relay cannot be adapted to the underlaying channels in each time slot. As a result $P_S^{(1)}(i)=P_S^{(1)}$,  $P_R^{(2)}(i)=P_R^{(2)}$, $P_S^{(3)}(i)=P_S^{(3)}$, and $P_R^{(3)}(i)=P_R^{(3)}$  hold $\forall i$.

The maximum achievable rate for this case can be found from (\ref{eq_op_AP}) by setting  $P_S^{(1)}(i)=P_S^{(1)}$,  $P_R^{(2)}(i)=P_R^{(2)}$, $P_S^{(3)}(i)=P_S^{(3)}$, and $P_R^{(3)}(i)=P_R^{(3)}$,    $\forall i$, in (\ref{eq_op_AP}). As  a result, we do not need to optimize in (\ref{eq_op_AP}) with respect to  these powers. Consequently, the constraints C4 and C5 in  (\ref{eq_op_AP}) can be removed. Thereby,   we get a new   optimization problem  for fixed-power allocation whose solutions are provided in the following theorem.

\begin{theorem}\label{Theo_FP}
The state selection variables $q_k(i)$, $\forall k,i$, maximizing the  achievable rate of the considered buffer-aided FD relay channel with SI for adaptive-rate transmission with fixed-power allocation (i.e., found as the solution of (\ref{eq_op_AP}) with  $P_S^{(1)}(i)=P_S^{(1)}$,  $P_R^{(2)}(i)=P_R^{(2)}$, $P_S^{(3)}(i)=P_S^{(3)}$, and $P_R^{(3)}(i)=P_R^{(3)}$, $\forall i$, and constraints C4-C5 removed) are given in (\ref{eq_scheme_AP}), where $\Lambda_1(i)$, $\Lambda_2(i)$,   and $\Lambda_3(i)$  are defined as
\vspace{-1mm}
\begin{align}
\Lambda_1(i) &= \mu \log_2\left(1 + P_S^{(1)}{\gamma _{SR}}(i)\right) ,\label{eq_FP_S11}\\
\Lambda_2(i) &= (1-\mu) {\log _2}\left(1 + P_R^{(2)}{\gamma _{RD}}(i)\right) ,\label{eq_FP_R12}\\
\Lambda_3(i) &= \mu  \log _2 \left( {1 + \frac{{P_S^{(3)}{\gamma _{SR}}(i)}}{{P_R^{(3)}{\gamma _{RR}}(i) + 1}}} \right)
 + (1-\mu){\log _2}\left(1 + P_R^{(3)}{\gamma _{RD}}(i)\right)\label{eq_FP_R23}.
\end{align}
In (\ref{eq_FP_S11})-(\ref{eq_FP_R23}), $\mu$ is a constant  found such that constraint C1 in (\ref{eq_op_AP}) holds. 
\end{theorem} 
\begin{IEEEproof}
Since the fixed-power allocation problem is a special case of (\ref{eq_op_AP}), when  $P_S^{(1)}(i)=P_S^{(1)}$,  $P_R^{(2)}(i)=P_R^{(2)}$, $P_S^{(3)}(i)=P_S^{(3)}$, and $P_R^{(3)}(i)=P_R^{(3)}$, $\forall i$, and when C4 and C5 are removed, we get the same solution as in (\ref{eq_AP_S1})-(\ref{eq_AP_R2}), but with $\zeta$ set to $\zeta=0$. This completes the proof.
\end{IEEEproof}

\begin{remark}
We note that in the extreme cases when  $\gamma_{RR}(i)\to\infty$ and  $\gamma_{RR}(i)\to 0$ hold, i.e., the communication schemes provided in Theorems~\ref{Theo_AP} and~\ref{Theo_FP} converge to the corresponding schemes in \cite{BA-relaying-adaptive-rate} and \cite{1435648}, respectively. In other words, in the extreme cases when we have infinite and zero SI,  the proposed buffer-aided schemes work as buffer-aided HD relaying and ideal FD relaying, respectively.
\end{remark}

\subsection{Practical Estimation of the Necessary Parameters}
The proposed state selection scheme, given in (\ref{eq_scheme_AP}), requires the computation of  $\Lambda_1(i)$, $\Lambda_2(i)$,  and $\Lambda_3(i)$ in each time slot. For the two proposed buffer-aided schemes,  these parameters can be computed at the FD relay with minimum possible channel state-information (CSI) acquisition overhead. Using $\Lambda_1(i)$, $\Lambda_2(i)$,  and $\Lambda_3(i)$, the relay can compute the optimal state selection variables  $q_1(i)$, $q_2(i)$, and $q_3(i)$ using (\ref{eq_scheme_AP}), and then feedback the optimal state to the source and the destination using two bits of feedback information. On the other hand, the computation of $\Lambda_1(i)$, $\Lambda_2(i)$,  and $\Lambda_3(i)$ at the relay requires full CSI of the S-R, R-D, and SI channels, as well as acquisition of the  constant $\mu$. Since $\mu$ is actually a Lagrange multiplier, in the following, we describe a method for estimating this constant   in real-time using only current instantaneous CSI by employing the gradient descent method \cite{Boyd_CO}.

In time slot $i$, we can recursively compute an estimate of the constant $\mu$, denoted by  $\mu_e(i)$, as 
\vspace{-1mm}
\begin{equation}\label{eq_mu_Rsr}
\mu_e(i) =\mu_e(i-1)  +\delta( i) \left[ \bar R_{RD}^e(i) - \bar R_{SR}^e(i)\right],
\end{equation}
where  $\bar R_{RD}^e(i)$ and $\bar R_{SR}^e(i)$ are real time estimates of $\bar R_{RD} $ and $\bar R_{SR} $, respectively, which can be calculated as
\vspace{-1mm}
\begin{align}
\bar R_{SR}^e(i)&=\frac{i-1}{i} \bar R_{SR}^e(i-1)+\frac{1}{i} C_{SR}(i),\label{eq_E_Rsr}\\
\bar R_{RD}^e(i)&=\frac{i-1}{i} \bar R_{RD}^e(i-1)+\frac{1}{i} C_{RD}(i).\label{eq_E_Rrd}
\end{align}
The values of $\bar R_{SR}^e(0)$ and $\bar R_{RD}^e(0)$ are initialized  to zero. Moreover, $\delta( i)$ is an adaptive step size which controls the speed of convergence of $\mu_e(i)$ to $\mu$, which can be some properly chosen monotonically decaying function of $i$ with $\delta( 1) < 1$.

For the buffer-aided relaying scheme with adaptive-power allocation, proposed in  Theorem~\ref{Theo_AP}, in addition to $\mu$, the constant $\zeta$ found in (\ref{eq_AP_S1})-(\ref{eq_AP_R2}) has to be acquired as well. This can be conducted in a similar manner as the  real-time estimation of $\mu$. In particular, in time slot $i$, we can recursively compute an estimate of the constant $\zeta$, denoted by  $\zeta_e(i)$, as 
\vspace{-1mm}
\begin{equation}\label{eq_zeta_Rsr}
\zeta_e(i)=\zeta_e(i-1) +\delta( i) \left[ \bar P_e(i) - P\right],
\end{equation}
where
\begin{align}
&\bar P_e(i)=\frac{i-1}{i} \bar P_e(i-1)
+\frac{1}{i}\left[q_1(i) P_S^{(1)}(i)+ q_2(i) P_R^{(2)}(i)+ q_3(i) \left( P_R^{(3)}(i) + P_S^{(3)}(i) \right)  \right],
\end{align}\label{eq_zeta_Rsr1}
where $P_S^{(1)}(i)$,  $P_S^{(3)}(i)$,  $P_R^{(2)}(i)$, and  $P_R^{(3)}(i)$ are given in Theorem~\ref{Theo_AP}. 

\section{Buffer-Aided Relaying with Fixed-Rate Transmission}\label{Sec-FR}

In this  section, we assume that the transmitting nodes, source and relay, do not have full CSI of their transmit links and/or have some other constraints that limit their ability to vary their transmission rates. As a result, we assume that when the source and the relay   are selected to transmit in a given time slot, they transmit with a fixed-rate   $R_0$. Moreover, we assume $P_S^{(1)}(i)=P_S^{(1)}$,  $P_R^{(2)}(i)=P_R^{(2)}$, $P_S^{(3)}(i)=P_S^{(3)}$, and $P_R^{(3)}(i)=P_R^{(3)}$, $\forall i$.
\vspace{-1mm}
\subsection{Derivation of the Proposed Buffer-Aided Relaying Scheme with Fixed-Rate Transmission}
 Due to the fixed-rate transmission,  outages may occur, i.e., in some time slots the data rate of the transmitted codeword, $R_0$, might be larger than the underlying AWGN channel capacity thereby leading to a received codeword which is undecodable at the corresponding receiver. To model the outages on the S-R  link, we define the following auxiliary binary variables
\vspace{-2mm}
\begin{align}
O_{SR}^{(1)}(i)=\left\{
\begin{array}{ll}
1 & \textrm{if }  \log_2\left(1 + P_S^{(1)}{\gamma _{SR}}(i)\right)\geq R_0\\
0 & \textrm{if }  \log_2\left(1 + P_S^{(1)}{\gamma _{SR}}(i)\right)< R_0,
\end{array}
\right.\label{eq_fr1}
\end{align}
\vspace{-4mm}
\begin{align}
O_{SR}^{(3)}(i)=\left\{
\setlength\arraycolsep{3pt}
\begin{array}{ll}
1 & \textrm{if } \log_2\left( 1 + \frac{P_S^{(3)}\gamma_{SR}(i)}{P_R^{(3)}\gamma_{RR}(i) + 1} \right) \geq R_0\\
0 & \textrm{if } \log_2\left( 1 + \frac{P_S^{(3)}\gamma_{SR}(i)}{P_R^{(3)}\gamma_{RR}(i) + 1} \right) < R_0.
\end{array}
\right. \label{eq_fr2}
\end{align}
Similarly, to model the outages on the R-D link, we define the following auxiliary binary variables 
\vspace{-2mm}
\begin{align}
O_{RD}^{(2)}(i)=\left\{
\setlength\arraycolsep{3pt}
\begin{array}{ll}
1 & \textrm{if }  \log_2\left(1 + P_R^{(2)}{\gamma _{RD}}(i)\right)\geq R_0\\
0 & \textrm{if }  \log_2\left(1 + P_R^{(2)}{\gamma _{RD}}(i)\right)< R_0,
\end{array}
\right.\label{eq_fr3}
\end{align}
\begin{align}
O_{RD}^{(3)}(i)=\left\{
\setlength\arraycolsep{3pt}
\begin{array}{ll}
1 & \textrm{if }  \log_2\left(1 + P_R^{(3)}{\gamma _{RD}}(i)\right)\geq R_0\\
0 & \textrm{if }  \log_2\left(1 + P_R^{(3)}{\gamma _{RD}}(i)\right)< R_0.
\end{array}
\right.\label{eq_fr3a}
\end{align}

Hence, a codeword transmitted by the source in time slot $i$ can be decoded correctly at the relay if and only if (iff) $q_1(i) O_{SR}^{(1)}  (i)+ q_3(i)O_{SR}^{(3)} (i)>0$ holds. Using $O_{SR}^{(1)}(i)$ and $O_{SR}^{(3)}(i)$, and the state selection variables $q_k(i)$, $\forall k$, we can define the data rate of the source in time slot $i$, $R_{SR}(i)$, as
\begin{align}
R_{SR}(i)= & R_0 \left[ q_1(i) O_{SR}^{(1)}  (i)+ q_3(i)O_{SR}^{(3)} (i) \right].\label{eq_2ac}
\end{align}  
In addition, we can obtain that a codeword transmitted by the relay in time slot $i$  can be decoded correctly at the destination iff $q_2(i) O_{RD}^{(2)}  (i)+ q_3(i)O_{RD}^{(3)} (i)>0$ holds. Similarly,  using $O_{RD}^{(2)}(i)$ and $O_{RD}^{(3)}(i)$, and the state selection variables $q_k(i)$, $\forall k$, we can define the data rate of the relay in time slot $i$, $R_{RD}(i)$, as
\begin{align}
R_{RD}(i) =& \min \left \{Q(i-1),R_0\left[ q_2(i) O_{RD}^{(2)}  (i)+ q_3(i)O_{RD}^{(3)} (i) \right] \right \},\label{eq_2bc}
\end{align}
where 
\begin{align}\label{eq_3cs}
Q(i)&=Q(i-1)+ R_{SR}(i)-R_{RD}(i).
\end{align}
The $\min \left \{\right\}$ in (\ref{eq_2bc}) is because the delay cannot transmit more information than the amount of information in its buffer $Q(i-1)$. Thereby,  the throughputs of the S-R and R-D channels during $N\to\infty$ time slots, again denoted by $\bar R_{SR}$ and $\bar R_{RD}$, can be obtained as
\begin{align}
\bar R_{SR} &=\lim_{N\to\infty}\frac 1N \sum_{i=1}^N R_0 \left[ q_1(i) O_{SR}^{(1)}  (i)+ q_3(i)O_{SR}^{(3)} (i) \right],\label{eq_fr4}\\
\bar R_{RD} &=\lim_{N\to\infty}\frac 1N  \sum_{i=1}^N \min \left\{ Q(i-1), 
R_0\left[ q_2(i) O_{RD}^{(2)}  (i)+ q_3(i)O_{RD}^{(3)} (i) \right] \right\}.\label{eq_fr5}
\end{align}

Our task in this section is to maximize the throughput of the considered relay channel with fixed-rate transmission given by (\ref{eq_fr5}). To this end, we use the following Lemma from    \cite[Theorem 1]{6408173}.
\begin{lemma}\label{lemma_2}
The  throughput   $\bar R_{RD}$, given by (\ref{eq_fr5}), is maximized when the following condition holds 
\begin{align}\label{eq_AD_FPaa} 
&\lim_{N\to\infty}\frac 1N \sum_{i=1}^N R_0 \left[ q_1(i) O_{SR}^{(1)}  (i)+ q_3(i)O_{SR}^{(3)} (i) \right]
\nonumber\\&\quad =\lim_{N\to\infty}\sum_{i=1}^N R_0 \left[ q_2(i) O_{RD}^{(2)}  (i)+ q_3(i)O_{RD}^{(3)} (i) \right].
\end{align}
Moreover, when condition (\ref{eq_AD_FPaa}) holds, the throughput, $\bar R_{RD}$, given by (\ref{eq_fr5}), simplifies to  
\begin{align}
\bar R_{RD}   = \lim_{N\to\infty}\sum_{i=1}^N R_0 \left[ q_2(i) O_{RD}^{(2)}  (i)+ q_3(i)O_{RD}^{(3)} (i) \right]\label{eq_4aaa}.
\end{align} 
\end{lemma}
\begin{IEEEproof}
See \cite[Theorem 1]{6408173} for the proof.
\end{IEEEproof}

Using Lemma~\ref{lemma_2}, we devise the following throughput maximization problem for the considered two-hop FD relay channel with SI and fixed-rate transmission
\begin{align}\label{eq_op_FR}
&\begin{array}{rl}
& {\underset{q_k(i),\forall i,k}{\textrm{Maximize: }}} \frac 1N  \sum_{i=1}^N R_0 \left[ q_2(i) O_{RD}^{(2)}  (i)+ q_3(i)O_{RD}^{(3)} (i) \right] \\
&{\rm{Subject\;\; to: }} \\
&\qquad{\rm C1:}\;   \frac 1N \sum_{i=1}^N R_0 \left[ q_1(i) O_{SR}^{(1)}  (i)+ q_3(i)O_{SR}^{(3)} (i) \right]
=\sum_{i=1}^N R_0 \left[ q_2(i) O_{RD}^{(2)}  (i)+ q_3(i)O_{RD}^{(3)} (i) \right] \\
&\qquad {\rm C2:}\; q_k(i)\in\{0,1\}, \textrm{ for } k=1,2,3  \\
\end{array} \nonumber\\
&\qquad\quad\,  {\rm C3:}\; q_1(i)+q_2(i)+q_3(i)\in\{0,1\}.
\end{align} 

In (\ref{eq_op_FR}), we maximize the throughput $\bar R_{RD}$, given by (\ref{eq_4aaa}), with respect to the state selection variables $ q_k(i)$, $\forall k$, when conditions (\ref{eq_1}) and (\ref{eq_AD_FPaa})  hold, and when $ q_k(i)$, $\forall k$, are binary. The solution of problem (\ref{eq_op_FR}) leads to the following theorem.

\begin{theorem}\label{Theo_FR}
The state selection variables  $ q_k(i)$, $\forall k$, maximizing the throughput of  the considered two-hop FD relay channel with SI and fixed-rate transmission, found as the solution of (\ref{eq_op_FR}), are given in (\ref{eq_scheme_AP}), 
where $\Lambda_1(i)$, $\Lambda_2(i)$, and $\Lambda_3(i)$  are defined as
\begin{align}
\Lambda_1(i) &= \mu O_{SR}^{(1)}(i),\label{eq_FR_1}\\
\Lambda_2(i) &= (1-\mu) O_{RD}^{(2)}(i) ,\label{eq_FR_2}\\
\Lambda_3(i) &= \mu O_{SR}^{(3)}(i)+ (1-\mu) O_{RD}^{(3)}(i).\label{eq_FR_3} 
\end{align}
In (\ref{eq_FR_1})-(\ref{eq_FR_3}), $\mu$ is a constant  found such that constraint C1 in (\ref{eq_op_FR}) holds. 
\end{theorem} 
\begin{IEEEproof}
Please refer to Appendix~\ref{app_FR} for the proof.
\end{IEEEproof}

\subsection{Practical Estimation of the Necessary Parameters}\label{SubSec-PracFR}
The fixed-rate transmission scheme, proposed in Theorem~\ref{Theo_FR}, requires the calculation of the parameter $\mu$. This parameter can be obtained theoretically by representing  constraint C1 in (\ref{eq_op_FR}) as  
\begin{align}\label{Pj_FR2}
E  \left\{ q_1(i) O_{SR}^{(1)}  (i)+ q_3(i)O_{SR}^{(3)} (i) \right\}
=E\left\{ q_2(i) O_{RD}^{(2)}  (i)+ q_3(i)O_{RD}^{(3)} (i) \right\}, 
\end{align}
where $E  \left\{ .\right\}$ denotes statistical expectation. Then, using the probability distribution functions (PDFs) of the S-R, R-D, and SI channels, the parameter $\mu$ can be found as the solution of (\ref{Pj_FR2}). Note that solving (\ref{Pj_FR2}) requires  knowledge of the PDFs of the channels. However, the scheme proposed in Theorem~\ref{Theo_FR} can still operate without any statistical knowledge in the following manner. We apply (\ref{eq_mu_Rsr}) to obtain an estimate of $\mu$ for time slot $i$, denoted by $\mu_e(i)$, where $C_{SR}(i)$ and $C_{RD}(i)$ in (\ref{eq_E_Rsr}) and (\ref{eq_E_Rrd}) are replaced by $q_1(i) O_{SR}^{(1)}(i)+ q_3(i)O_{SR}^{(3)}(i)$ and $q_2(i) O_{RD}^{(2)}(i)+ q_3(i)O_{RD}^{(3)}(i)$, respectively.   We now let $\mu_e(i)$ to take any value in the range $[ 0,1 ]$. Then, when two state selection variables both assume the value one, according to (\ref{eq_scheme_AP}), we select one at random with equal probability to assume the value one and set the other state selection variable to zero. For this scheme, we choose $\delta(i)$ to vary with $i$ only during the first several time slots, and then we set it to a constant for the remaining time slots.
 
\vspace{-2mm}
\section{Simulation and Numerical Results}\label{Sec-Num}
In this section, we evaluate the performance of the proposed buffer-aided schemes with adaptive reception-transmission at the FD relay for the two-hop FD   relay channel with SI, and compare it to the performance of several benchmark schemes. To this end, we first define the system parameters, then introduce the  benchmark schemes and a delay-constrained buffer-aided relaying scheme, and finally present the numerical results.

\subsection{System Parameters}
For the presented numerical results, the mean of the channel gains of the S-R and R-D links are calculated using the standard path loss model as
\begin{eqnarray}
E\{|h_L(i)|^2\} = \left(\frac{c}{{4\pi {f_c}}}\right)^2d_L^{ - \alpha }\ \textrm{, for } L\in\{\textrm{S-R},\textrm{R-D}\},
\end{eqnarray}
where $c$ is the speed of light, $f_c$ is the carrier frequency, $d_L$ is the distance between the transmitter and the receiver of link $L$, and $\alpha$ is the path loss exponent. In this section, we set $\alpha=3$, $f_c=2.4$ GHz, and $d_L=500$ m for $L\in\{\textrm{S-R},\textrm{R-D}\}$. Moreover, we assume that the transmit bandwidth is   $200$ kHz, and the noise power per Hz is    $-170$ dBm. Hence,  the  total noise power for $200$ kHz is obtained as $-117$ dBm. On the other hand, the value of $E\{|h_{RR}(i)|^2\}$  is set to range between $-100$ dB to $-140$ dB.  Note that $E\{|h_{RR}(i)|^2\}$ can be considered as the SI suppression factor of the corresponding SI suppression scheme.

\subsection{Benchmark Schemes}\label{Sec-BenSch}
In the following, we introduce three benchmark schemes which will be used for benchmarking the proposed buffer-aided relaying schemes. For the benchmark schemes $P_S(i)=P_S$ and $P_R(i)=P_R$, $\forall i$, is assumed.

\textit{Benchmark Scheme 1 (Buffer-aided HD relaying with adaptive reception-transmission):} The  achievable rate of employing an HD relay and using the buffer-aided HD relaying scheme with adaptive reception-transmission  proposed in \cite{BA-relaying-adaptive-rate}, is given in [6, Section \ref{Sec-BAR}-D].

\textit{Benchmark Scheme 2 (Conventional FD relaying with a buffer):} In the conventional FD relaying scheme, the FD relay \textit{simultaneously transmits and receives} during all time slots with $P_S=tP$ and $P_R=(1-t)P$.  Because there is a buffer at the FD relay, the received information can be stored and transmitted   in future time slots. As a result  the achieved data rate during $N\to\infty$ time slots is given by 
\begin{align}
R_{FD,2}=& \min \bigg \{ \mathop {\lim }\limits_{N \to \infty } \frac{1}{N}\sum\limits_{i = 1}^N {{{\log }_2}} \left( {1 + \frac{{tP{\gamma _{SR}}(i)}}{{(1 - t)P{\gamma _{RR}}(i) + 1}}} \right),\nonumber\\
&\mathop {\lim }\limits_{N \to \infty } \frac{1}{N}\sum\limits_{i = 1}^N {{{\log }_2}} \left(1 + (1 - t)P{\gamma _{RD}}(i) \right)\bigg\}.
\end{align}
We note that the conventional FD relaying scheme without a buffer achieves a worse performance than the conventional FD relaying scheme with a buffer. As a result, this scheme is not used as a benchmark.

\textit{Benchmark Scheme 3 (FD relaying with an ideal FD relay without SI):}  
This is identical to the Benchmark Scheme 2, except that ${\gamma _{RR}}(i)$ is set to zero, i.e, to $-\infty$ dB.

\subsection{Buffer-Aided Relaying Schemes for Delay-Constrained Transmission}\label{Sec-DC}
The proposed scheme in (\ref{eq_scheme_AP}) gives the maximum achievable rate and the maximum throughput, however, it cannot fix the delay to a desired level. In the following, similar to  \cite{7084188}, we propose a scheme for the state selection variables $q_k(i)$, $\forall k$, which holds the delay at a desired level.

The average delay of the considered network is given by Little's Law, as
\begin{align}\label{DC_ADelay}
E\{T(i)\}=\lim_{N\to\infty} \frac{  \sum_{i=1}^N Q(i) }{ \sum_{i=1}^N  R_{SR}(i)},
\end{align}
where $R_{SR}(i)$ and $Q(i)$ are the transmission rate of the source and the queue length in time slot $i$, respectively. The rate $R_{SR}(i)$ is defined in (\ref{eq_2a}) for adaptive-rate transmission and in (\ref{eq_2ac}) for fixed-rate transmission. Moreover, $Q(i)$ is defined in (\ref{eq_3c}) for adaptive-rate transmission and in (\ref{eq_3cs}) for fixed-rate transmission.
 
For the proposed delay-constrained scheme, we continue to use the general state selection scheme in (\ref{eq_scheme_AP}), where  $\Lambda_1(i)$, $\Lambda_2(i)$,   and $\Lambda_3(i)$, are calculated  using (\ref{eq_AP_S1})-(\ref{eq_AP_R2}) for adaptive-rate transmission with adaptive-power allocation,  using (\ref{eq_FP_S11})-(\ref{eq_FP_R23}) for adaptive-rate transmission with fixed-power allocation, and using (\ref{eq_FR_1})-(\ref{eq_FR_3}) for fixed-rate transmission. Note that $\Lambda_1(i)$, $\Lambda_2(i)$,   and $\Lambda_3(i)$ in  (\ref{eq_AP_S1})-(\ref{eq_AP_R2}), (\ref{eq_FP_S11})-(\ref{eq_FP_R23}), and  (\ref{eq_FR_1})-(\ref{eq_FR_3}) are all function of the parameter $\mu$. In contrary to the proposed buffer-aided schemes in Theorems~\ref{Theo_AP}-\ref{Theo_FR}, where $\mu$ is found to satisfy the constant C1 in (\ref{eq_op_AP}) and (\ref{eq_op_FR}), in the buffer-aided relaying scheme for delay-constrained transmission, we use $\mu$ to ensure that the system achieves a desired average delay. To this end, $\mu$ is calculated as follows. Let us define $T_0$ as the desired average delay constraint of the considered relay network. Then, in time slot $i$, we can recursively compute an estimate of the constant $\mu$, denoted by  $\mu_e(i)$, as 
\begin{equation}\label{eq_mu_Rsr_DC}
\mu_e(i) =\mu_e(i-1)  +\delta( i) \left[ T_0 - \frac{Q(i)}{\bar R_{SR}^e(i)} \right],
\end{equation}
where $Q(i)$ is the queue length, defined in (\ref{eq_3c}) and (\ref{eq_3cs}) for adaptive-rate transmission with fixed- and adaptive-power allocation schemes and fixed-rate transmission scheme, respectively. Furthermore, $\bar R_{SR}^e(i)$ is a real time estimate of $\bar R_{SR} $, calculated using (\ref{eq_E_Rsr}), and initialized  to zero for $i=0$, i.e., $\bar R_{SR}^e(0)=0$. Moreover, $\delta( i)$ is an adaptive step function, which can be chosen to be a properly monotonically decaying function of $i$ with $\delta( 1) < 1$.

\subsection{Numerical Results}
  All of the presented results in this section are generated for Rayleigh fading by numerical evaluation of the derived results and are confirmed by Monte Carlo simulations.

\begin{figure}[t]
\vspace*{-2mm}
\centering\includegraphics[width=5in]{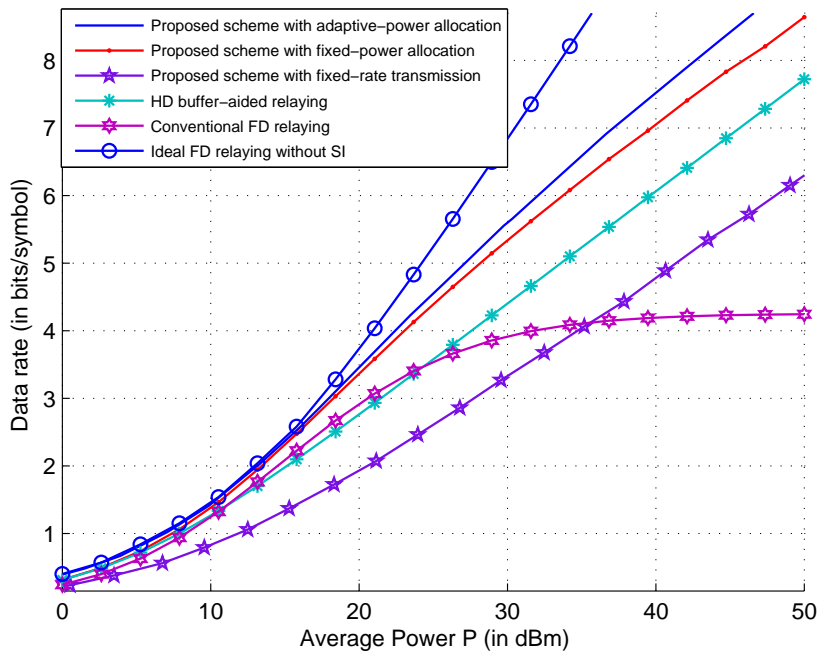}
\caption{Data rate vs. average consumed power of the proposed schemes and the benchmark schemes.}
\label{Benchmarks_TP_Power}
\vspace*{-3mm}
\end{figure}

In Fig.~\ref{Benchmarks_TP_Power}, the rates achieved using the proposed FD buffer-aided scheme for adaptive-rate transmission with adaptive-power allocation, adaptive-rate transmission with fixed-power allocation,  and fixed-rate transmission are compared with the benchmark schemes, defined in Section \ref{Sec-BenSch}, as a function of the average consumed power, $P$. The average gain of the SI channel in this numerical example is set to ${|h_{RR}(i)|}^2=-133\textrm{ dB}$. For the proposed scheme with adaptive-rate transmission with fixed-power allocation  and fixed-rate transmission we set $P_S^{(1)}(i)=P$, $P_R^{(2)}(i)=P$, $P_S^{(3)}(i)=tP$ and $P_R^{(3)}(i)=(1-t)P$, where $t=0.5$, $\forall i$. Moreover, for the proposed scheme with fixed-rate transmission the value of $R_0$ is optimized numerically, for a given average power $P$, such that the throughput is maximized. As can be seen clearly from Fig.~\ref{Benchmarks_TP_Power}, the proposed buffer-aided schemes with and without power allocation achieve substantial gains compared to the benchmark schemes.   For example, the proposed adaptive-rate transmission with adaptive-power allocation scheme has a power gain of   about $2\textrm{ dB}$, $6\textrm{ dB}$ and $9\textrm{ dB}$ compared to buffer-aided HD relaying for rates of 2, 5, and 7 bits per symbol, respectively. Moreover, as can be seen from Fig.~\ref{Benchmarks_TP_Power}, the performance of the conventional FD relaying scheme, i.e, Benchmark Scheme 2, is very poor. In fact, even the proposed fixed-rate transmission scheme outperforms the conventional FD scheme for $P>35$ dBm. This numerical result clearly shows the significant gains that can be achieved with the proposed buffer-aided schemes compared to  previous schemes available in the literature.

\begin{figure}[t]
\vspace*{-1mm}
\centering\includegraphics[width=5in]{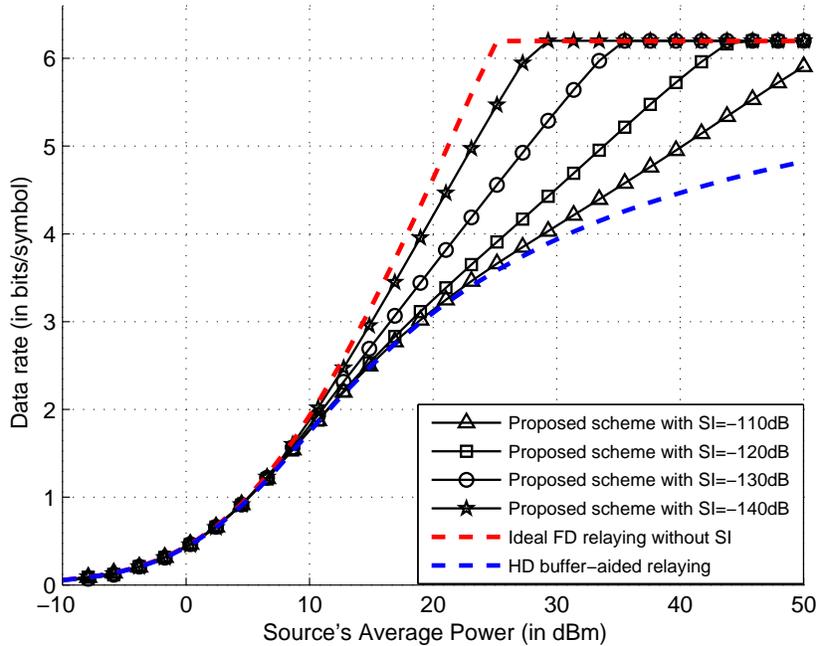}
\caption{Data rate vs. the average consumed power at the source node of the proposed buffer-aided scheme for adaptive-rate transmission with fixed-power allocation and the benchmark schemes when the average power at the relay is set to  25 dBm.}
\label{Benchmark_Power_Fixed_R}
\vspace*{-3mm}
\end{figure}

In Fig.~\ref{Benchmark_Power_Fixed_R}, we compare the achievable rates of the proposed buffer-aided scheme for adaptive-rate transmission with fixed-power allocation, with the capacity of the ideal two-hop FD relay channel without SI and with the rate achieved by buffer-aided HD relaying as a function of the average transmit power at the source node, where $P_S^{(1)}=P_S^{(3)}=P_S$ is adopted, for different values of the SI. In this example, the power of the relay, $P_R^{(2)}=P_R^{(3)}=P_R$, is set to 25 dBm. This example models an HD base-station which can vary its average power $P_S$, that is helped by an FD relay with fixed average power $P_R=25\textrm{ dBm}$ to transmit information to a destination. Since the transmitted power at the relay node is fixed, the maximum possible data rate on the R-D channel is around 6.2 bits per symbol. As can be observed from Fig.~\ref{Benchmark_Power_Fixed_R}, the performance of the proposed FD buffer-aided scheme is considerably larger than the performance of buffer-aided HD relaying when the transmit power at the source is larger than 25 dBm. For example, for 5 bits per symbol, the power gains are approximately 30 dB, 25 dB, 20 dB, and 15 dB compared to HD relaying for SI values of -140 dB, -130 dB, -120 dB, and -110 dB, respectively. We can see that the proposed scheme works much better (i.e., 15 dB gain) than the best known HD buffer-aided relaying scheme for a fair SI level of $-110\textrm{ dB}$, which shows that indeed it is beneficial for FD relays to be employed around HD base stations in order to increase their performance significantly.

\begin{figure}[t]
\centering\includegraphics[width=5in]{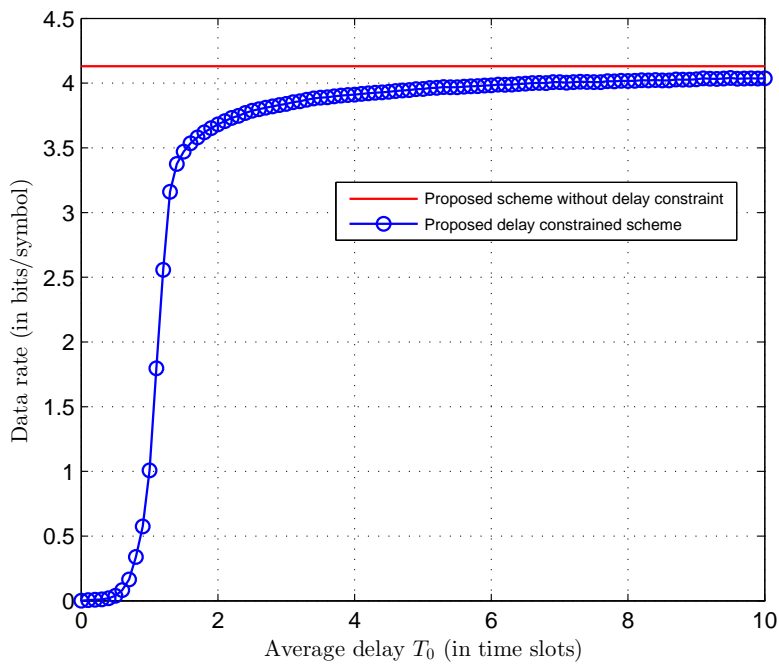}
\caption{Average delay until time slot $i$ vs. time slot $i$, for proposed scheme with delay-constrained transmission compared to fixed desired value, $T_0$.}
\label{Rate_Vs_Delay}
\end{figure}

In Fig.~\ref{Rate_Vs_Delay}, we demonstrate the achievable rate of the proposed delay-constrained buffer-aided scheme as a function of the average desired delay, $T_0$, and compare it with the maximum achievable rate obtained with the scheme in Theorem~\ref{Theo_FP}, which does not fix the delay. For this numerical example, we set $P_S^{(1)}(i)=24\textrm{ dBm}$, $P_R^{(2)}(i)=24\textrm{ dBm}$, $P_S^{(3)}(i)=21\textrm{ dBm}$ and $P_R^{(3)}(i)=21\textrm{ dBm}$, $\forall i$. We can see from Fig.~\ref{Rate_Vs_Delay} that by increasing the average delay, $T_0$, both data rates converge fast. In fact, for a delay of 3 time slots, there is only $7\%$ loss compared to the maximum rate. This shows that the proposed delay-constrained buffer-aided scheme achieves rate close to the maximum possible rate for a very small delay.

\begin{figure}[t]
\centering\includegraphics[width=5in]{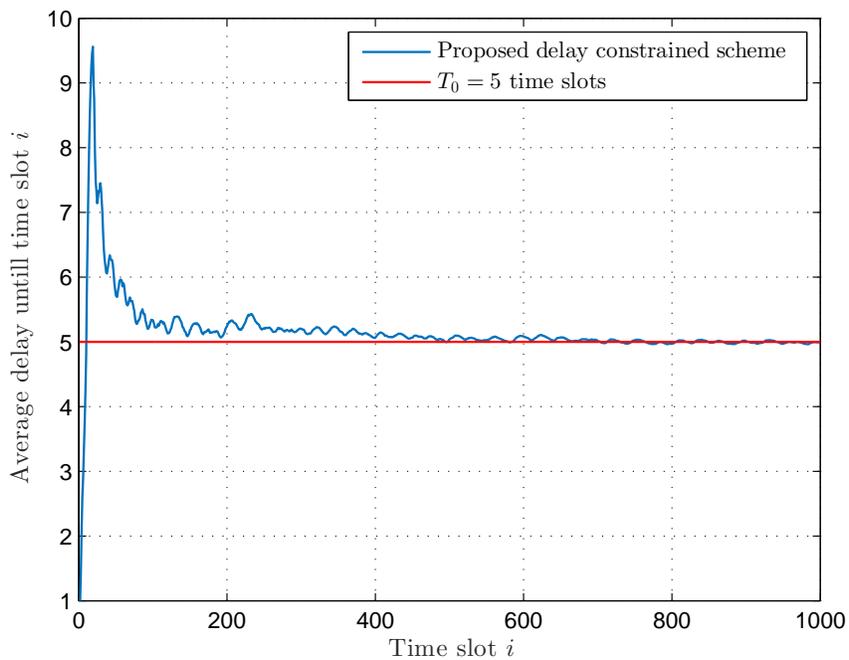}
\caption{Average delay until time slot $i$ vs. time slot $i$, for proposed scheme with delay-constrained transmission compared to fixed desired value, $T_0$.}
\label{FixedPower_Delay_Time}
\end{figure}

In Fig.~\ref{FixedPower_Delay_Time}, we plot the average delay of the proposed delay-constrained scheme
until time slot $i$, for the case when $T_0=5$ time slots, as a function of time slot $i$. Fig.~\ref{FixedPower_Delay_Time} reveals that the average delay until time slot $i$ converges very fast to $T_0$ by increasing $i$. Moreover, when the average delay converges to its desired level, it has relatively small fluctuations around it. This shows that the proposed delay-constrained scheme is very effective in reaching the desired level of delay.

\vspace{-1mm}
\section{Conclusion}\label{Sec-Conc}
In this paper, we proposed three buffer-aided relaying schemes with adaptive reception-transmission at the FD relay for the two-hop FD relay channel with SI for the cases of adaptive-rate transmission with adaptive-power allocation,  adaptive-rate transmission with fixed-power allocation,  and   fixed-rate transmission, respectively. The proposed schemes significantly improve the over-all performance by optimally selecting the FD relay to either receive, transmit, or simultaneously receive and transmit in a given time slot based on the qualities of the receiving, transmitting, and SI channels. Also, we proposed a relatively fast and practical buffer-aided scheme that holds the delay around a desirable value. Our numerical results have shown that significant performance gains are achieved using the proposed buffer-aided relaying schemes compared to   conventional FD relaying, where the FD relay is forced to always simultaneously receive and transmit, and to   buffer-aided HD relaying, where the HD relay  cannot simultaneously receive and transmit. This means that buffer-aided relaying should become an integral part of future FD relaying systems, i.e., i.e., that FD relaying systems without buffer-aided relaying miss-out on significant performance gains.

 
\appendix 
\subsection{Proof of Theorem~\ref{Theo_AP}}\label{app_AP}

We relax constraints C2 and C3 such that $0\leq q_k(i)\leq 1$, $\forall k$, and $0\leq q_1(i)+q_2(i)+q_3(i)\leq 1$, and ignore constraint C5. Then, we use the Lagrangian method for solving this optimization problem. Thereby, with some simplification, we can obtain the Lagrangian function, ${\cal L}$, as
\begin{align}\label{eq_op_AP_append}
&\begin{array}{ll}
{\cal L} &= -q_2(i) {\log _2}\left(1 + P_R^{(2)}(i){\gamma _{RD}}(i)\right) - q_3(i) {\log _2}\left(1 + P_R^{(3)}(i){\gamma _{RD}}(i)\right) \\
&- \mu \bigg[- q_2(i) {\log _2}\left(1 + P_R^{(2)}(i){\gamma _{RD}}(i)\right) - q_3(i) {\log _2}\left(1 + P_R^{(3)}(i){\gamma _{RD}}(i)\right)\\
&+ q_1(i) {\log _2}\left(1 + P_S^{(1)}(i){\gamma _{SR}}(i)\right) + q_3(i){\log _2}\left( {1 + \frac{{P_S^{(3)}(i) {\gamma _{SR}}(i)}}{{P_R^{(3)}(i) {\gamma _{RR}}(i) + 1}}} \right) \bigg]\\
& \zeta \left(\:q_1(i) P_S^{(1)}(i)+ q_3(i) P_S^{(3)}(i) + q_2(i) P_R^{(2)}(i) + q_3(i) P_R^{(3)}(i)\:\right) \\
&- {\lambda _1(i)}{q_1(i)} - {\lambda _2(i)}(1 - {q_1(i)})- {\lambda _3(i)}{q_2(i)} \\
&- {\lambda _4(i)}(1 - {q_2(i)}) - {\lambda _5(i)}{q_3(i)} - {\lambda _6(i)}(1 - {q_3(i)})\\
\end{array}\nonumber\\[-0.5em]
& \quad\,\,\,\,-{\lambda _7(i)}({q_1(i)} + {q_2(i)} + {q_3(i)}) - {\lambda _8(i)}(1 - ({q_1(i)} + {q_2(i)} + {q_3(i)})), 
\end{align}
where $\mu$, $\zeta\geq0$ and $\lambda_{k}(i)\geq0$, are the Lagrangian multipliers. By differentiating ${\cal L}$ with respect to $P_S^{(k)}(i)$ and $P_R^{(k)}(i)$, $\forall k$, and then setting the result to zero, we obtain
\begin{align}
\frac{{d{\cal L}}}{{d{P_S^{(1)}(i)}}} &= \frac{{\mu {q_1(i)}{\gamma _{SR}}(i)}}{{\ln(2){\mkern 1mu} (1 + {P_S^{(1)}(i)}{\gamma _{SR}}(i))}}-\zeta {q_1(i)} =0,\label{eq_AP_power_eq1}\\
\frac{{d\cal L}}{{d{P_R^{(2)}(i)}}} &= \frac{{(1 - \mu ){q_2(i)}{\gamma _{RD}}(i)}}{{\ln(2){\mkern 1mu} (1 + {P_R^{(2)}(i)}{\gamma _{RD}}(i))}}  - \zeta {q_2}(i) = 0,\label{eq_AP_power_eq2}\\
\frac{{d\cal L}}{{d{P_S^{(3)}(i)}}} &=\frac{{\mu {q_3(i)}{\gamma _{SR}}(i)}}{{\ln(2){\mkern 1mu} (1 + {P_R^{(3)}(i)}{\gamma _{RR}}(i) + {P_S^{(3)}(i)}{\gamma _{SR}}(i))}} - \zeta {q_3}(i) = 0,\label{eq_AP_power_eq3}\\
\frac{{d\cal L}}{{d{P_R^{(3)}(i)}}} &= \frac{{(1 - \mu ){q_3}(i){\gamma _{RD}}(i)}}{{\ln(2){\mkern 1mu} (1 + {P_R^{(3)}(i)}{\gamma _{RD}}(i))}} - \zeta {q_3}(i)\nonumber\\
&- \frac{{\mu {q_3}(i){\gamma _{RR}}(i){\gamma _{SR}}(i){P_S^{(3)}(i)}}}{{\ln(2){\mkern 1mu} (1 + {P_R^{(3)}(i)}{\gamma _{RR}}(i)){\mkern 1mu} (1 + {P_R^{(3)}(i)}{\gamma _{RR}}(i) + {P_S^{(3)}(i)}{\gamma _{SR}}(i))}}= 0.\label{eq_AP_power_eq4}
\end{align}

Now, we calculate $P_S^{(k)}(i)$ and $P_R^{(k)}(i)$, $\forall k$, based on equation (\ref{eq_AP_power_eq1})-(\ref{eq_AP_power_eq4}), and the following three different available states.

• $q_{1}(i)=1$: Since $q_{1}(i)=1$, we set $q_{2}(i)=0$ and $q_{3}(i)=0$. As a result (\ref{eq_AP_power_eq1}) becomes
\begin{equation}
\frac{{d{\cal L}}}{{d{P_S^{(1)}(i)}}} = \frac{{\mu {\gamma _{SR}}(i)}}{{\ln(2){\mkern 1mu} (1 + {P_S^{(1)}(i)}{\gamma _{SR}}(i))}}-\zeta =0.\label{eq_AP_power_PS1}
\end{equation}
Solving (\ref{eq_AP_power_PS1}), we can obtain $P_S^{(1)}(i)$ as in (\ref{eq_scheme_AP_PS1}).

• $q_{2}(i)=1$: Since $q_{2}(i)=1$, we set $q_{1}(i)=0$ and $q_{3}(i)=0$. As a result, (\ref{eq_AP_power_eq2}) becomes
\begin{equation}
\frac{{d\cal L}}{{d{P_R^{(2)}(i)}}} = \frac{{(1 - \mu ){\gamma _{RD}}(i)}}{{\ln(2){\mkern 1mu} (1 + {P_R^{(2)}(i)}{\gamma _{RD}}(i))}}  - \zeta = 0.\label{eq_AP_power_PR2}
\end{equation}
Solving (\ref{eq_AP_power_PR2}), we can obtain $P_R^{(2)}(i)$, as in (\ref{eq_scheme_AP_PR2}).

• $q_{3}(i)=1$: Since $q_{3}(i)=1$, we set $q_{1}(i)=0$ and $q_{2}(i)=0$. As a result, (\ref{eq_AP_power_eq3}) and (\ref{eq_AP_power_eq4}) simplify to (\ref{eq_AP_PR31}) and (\ref{eq_AP_PR32}), respectively. By solving (\ref{eq_AP_PR31}) and (\ref{eq_AP_PR32}), we can obtain $P_S^{(3)}(i)$ and $P_R^{(3)}(i)$. We note that, although in this case there is a closed form solution for $P_S^{(3)}(i)$ and $P_R^{(3)}(i)$, since the solution is very long, we have decided not to show it in this paper.

The Lagrangian function, ${\cal L}$, given by  (\ref{eq_op_AP_append}) is bounded below if and only if 
\begin{align}
&-\mu \log_2\left(1 + {P_S^{(1)}(i)}{\gamma _{SR}}(i)\right) +\zeta {P_S^{(1)}(i)}-\lambda _1(i)+\lambda _2(i)-\lambda _7(i)+\lambda _8(i)=0,\label{EQ_Dual_1}\\
&-(1 - \mu )\log_2\left(1 + {P_R^{(2)}(i)}{\gamma _{RD}}(i)\right) + \zeta {P_R^{(2)}(i)}-\lambda _3(i)+\lambda _4(i)-\lambda _7(i)+\lambda _8(i)=0,\label{EQ_Dual_2}\\
&-(1 - \mu )\log_2\left(1 + {P_R^{(3)}(i)}{\gamma _{RD}}(i)\right) - \mu \log_2\left(1 + \frac{{{P_S^{(3)}(i)}{\gamma _{SR}}(i)}}{{{P_R^{(3)}(i)}{\gamma _{RR}}(i) + 1}}\right) \nonumber\\
 &+\zeta {P_S^{(3)}(i)} + \zeta {P_R^{(3)}(i)}-\lambda _5(i)+\lambda _6(i)-\lambda _7(i)+\lambda _8(i)=0.\label{EQ_Dual_3}
\end{align}
We define $-\lambda_{7}(i)+\lambda_{8}(i)\triangleq \beta(i)$, and find the system selection schemes for the three different available cases as follows:

• $q_{1}(i)=1$: Since $q_{1}(i)=1$, we set $q_{2}(i)=0$ and $q_{3}(i)=0$. As a result, we have $\lambda_1(i)=0$, $\lambda_4(i)=0$  and $\lambda_6(i)=0$ (by complementary slackness in KKT condition), which using them we can rewrite  (\ref{EQ_Dual_1}), (\ref{EQ_Dual_2}), and (\ref{EQ_Dual_3}),  as  
\begin{align}\label{eq_AP_power_q1}
 &\quad\,\,\mu \log_2\left(1 + {P_S^{(1)}(i)}{\gamma _{SR}}(i)\right) - \zeta {P_S^{(1)}(i)} - \beta(i)  > 0,\\
&\begin{array}{ll}
& (1 - \mu )\log_2\left(1 + {P_R^{(2)}(i)}{\gamma _{RD}}(i)\right) - \zeta {P_R^{(2)}(i)} - \beta(i)  < 0,\\
 &\textrm{And} \:(1 - \mu )\log_2\left(1 + {P_R^{(3)}(i)}{\gamma _{RD}}(i)\right) + \mu \log_2\left(1 + \frac{{{P_S^{(3)}(i)}{\gamma _{SR}}(i)}}{{{P_R^{(3)}(i)}{\gamma _{RR}}(i) + 1}}\right) \\
 &\qquad- \zeta {P_S^{(3)}(i)} - \zeta {P_R^{(3)}(i)} - \beta(i)<0,
\end{array}\nonumber
\end{align}
respectively.

• $q_{2}(i)=1$: Since $q_{2}(i)=1$, we set $q_{1}(i)=0$ and $q_{3}(i)=0$. As a result, we have $\lambda_2(i)=0$, $\lambda_3(i)=0$  and $\lambda_6(i)=0$, which using them we can rewrite  (\ref{EQ_Dual_1}), (\ref{EQ_Dual_2}), and (\ref{EQ_Dual_3}),  as  
\begin{align}\label{eq_AP_power_q2}
 &\quad\,\,\mu \log_2\left(1 + {P_S^{(1)}(i)}{\gamma _{SR}}(i)\right) - \zeta {P_S^{(1)}(i)} - \beta(i)  < 0,\\
&\begin{array}{ll}
 &(1 - \mu )\log_2\left(1 + {P_R^{(2)}(i)}{\gamma _{RD}}(i)\right) - \zeta {P_R^{(2)}(i)} - \beta(i)  > 0,\\
 &\textrm{And} \:(1 - \mu )\log_2\left(1 + {P_R^{(3)}(i)}{\gamma _{RD}}(i)\right) + \mu \log_2\left(1 + \frac{{{P_S^{(3)}(i)}{\gamma _{SR}}(i)}}{{{P_R^{(3)}(i)}{\gamma _{RR}}(i) + 1}}\right) \\
 &\qquad- \zeta {P_S^{(3)}(i)} - \zeta {P_R^{(3)}(i)} - \beta(i)<0,
\end{array}\nonumber
\end{align}
respectively.

• $q_{3}(i)=1$: Since $q_{3}(i)=1$, we set $q_{1}(i)=0$ and $q_{2}(i)=0$. As a result, we have $\lambda_2(i)=0$, $\lambda_4(i)=0$  and $\lambda_5(i)=0$, which using them we can rewrite  (\ref{EQ_Dual_1}), (\ref{EQ_Dual_2}), and (\ref{EQ_Dual_3}),  as  
\begin{align}\label{eq_AP_power_q3}
 &\quad\,\,\mu \log_2\left(1 + {P_S^{(1)}(i)}{\gamma _{SR}}(i)\right) - \zeta {P_S^{(1)}(i)} - \beta(i)  < 0,\\
&\begin{array}{ll}
 &(1 - \mu )\log_2\left(1 + {P_R^{(2)}(i)}{\gamma _{RD}}(i)\right) - \zeta {P_R^{(2)}(i)} - \beta(i)  < 0,\\
 &\textrm{And} \:(1 - \mu )\log_2\left(1 + {P_R^{(3)}(i)}{\gamma _{RD}}(i)\right) + \mu \log_2\left(1 + \frac{{{P_S^{(3)}(i)}{\gamma _{SR}}(i)}}{{{P_R^{(3)}(i)}{\gamma _{RR}}(i) + 1}}\right) \\
 &\qquad- \zeta {P_S^{(3)}(i)} - \zeta {P_R^{(3)}(i)} - \beta(i)>0,
\end{array}\nonumber
\end{align}
respectively.

By substituting the corresponding terms in (\ref{eq_AP_power_q1}), (\ref{eq_AP_power_q2}), and (\ref{eq_AP_power_q3}) by  (\ref{eq_AP_S1}), (\ref{eq_AP_R1}), and (\ref{eq_AP_R2}), we can derive the optimal state selection scheme in Theorem~\ref{Theo_AP}. This completes the proof.

\subsection{Proof of Theorem~\ref{Theo_FR}}\label{app_FR}
We use the Lagrangian method for solving (\ref{eq_op_FR}). With some simplification, we can obtain the Lagrangian function as
\begin{align}
{\cal L}=&-q_1(i)\mu R_0 O_{SR}^{(1)}(i)- q_2(i)(1-\mu) R_0 O_{RD}^{(2)}(i)- q_3(i) \left[ \mu R_0 O_{SR}^{(3)}(i)+(1-\mu) R_0 O_{RD}^{(3)}(i) \right]\nonumber\\
&- q_1(i) {\log _2}\left(1 + P_S^{(1)}(i){\gamma _{SR}}(i)\right) - q_3(i){\log _2}\left( {1 + \frac{{P_S^{(3)}(i) {\gamma _{SR}}(i)}}{{P_R^{(3)}(i) {\gamma _{RR}}(i) + 1}}} \right) \bigg]\nonumber\\
&- {\lambda _1(i)}{q_1(i)} - {\lambda _2(i)}(1 - {q_1(i)})- {\lambda _3(i)}{q_2(i)} \nonumber\\
&- {\lambda _4(i)}(1 - {q_2(i)}) - {\lambda _5(i)}{q_3(i)} - {\lambda _6(i)}(1 - {q_3(i)})\nonumber\\
&-{\lambda _7(i)}({q_1(i)} + {q_2(i)} + {q_3(i)}) - {\lambda _8(i)}(1 - ({q_1(i)} + {q_2(i)} + {q_3(i)})),\label{eq_op_FR_append} 
\end{align}
where $\mu\geq0$ and $\lambda_{k}(i)\geq0$ are the Lagrangian multipliers. Similar to Appendix~\ref{app_AP}, we can find  $ q_k(i)$, $\forall k$, which maximize (\ref{eq_op_FR_append}) as follows. To this end, in the Lagrangian function, ${\cal L}$, given by  (\ref{eq_op_FR_append}), we define $-\lambda_{7}(i)+\lambda_{8}(i)\triangleq \beta(i)$, and find the system selection schemes for the three different available cases as follows:

• $q_{1}(i)=1$: Since $q_{1}(i)=1$, we set $q_{2}(i)=0$ and $q_{3}(i)=0$. As a result, the conditions which maximize (\ref{eq_op_FR_append}) in this case are
\begin{align}\label{eq_FR_power_q1}
 &\mu R_0 O_{SR}^{(1)}(i) - \beta(i)  > 0,\nonumber\\
 &\textrm{And} \:(1-\mu) R_0 O_{RD}^{(2)}(i) - \beta(i)  < 0,\nonumber\\
 &\textrm{And} \:\left[ \mu R_0 O_{SR}^{(3)}(i)+(1-\mu) R_0 O_{RD}^{(3)}(i) \right] - \beta(i)<0.
\end{align}
• $q_{2}(i)=1$: Since $q_{2}(i)=1$, we set $q_{1}(i)=0$ and $q_{3}(i)=0$. For maximizing (\ref{eq_op_FR_append}), the following conditions must be held
\begin{align}\label{eq_FR_power_q2}
 &\mu R_0 O_{SR}^{(1)}(i) - \beta(i)  < 0,\nonumber\\
 &\textrm{And} \:(1-\mu) R_0 O_{RD}^{(2)}(i) - \beta(i)  > 0,\nonumber\\
 &\textrm{And} \:\left[ \mu R_0 O_{SR}^{(3)}(i)+(1-\mu) R_0 O_{RD}^{(3)}(i) \right] - \beta(i)<0.
\end{align}
• $q_{3}(i)=1$: Since $q_{3}(i)=1$, we set $q_{1}(i)=0$ and $q_{2}(i)=0$. Finally, the conditions which maximize (\ref{eq_op_FR_append}), in this case are
\begin{align}\label{eq_FR_power_q3}
 &\mu R_0 O_{SR}^{(1)}(i) - \beta(i)  < 0,\nonumber\\
 &\textrm{And} \:(1-\mu) R_0 O_{RD}^{(2)}(i) - \beta(i)  < 0,\nonumber\\
 &\textrm{And} \:\left[ \mu R_0 O_{SR}^{(3)}(i)+(1-\mu) R_0 O_{RD}^{(3)}(i) \right] - \beta(i)>0.
\end{align}
By substituting the corresponding terms in (\ref{eq_FR_power_q1}), (\ref{eq_FR_power_q2}), and (\ref{eq_FR_power_q3}) by  (\ref{eq_FR_1}), (\ref{eq_FR_2}), and (\ref{eq_FR_3}), and dividing by the common term $R_0$,  we obtain the optimal state selection scheme in Theorem~\ref{Theo_FR}. This completes the proof. 
 
\bibliography{litdab}
\bibliographystyle{IEEEtran}
\end{document}